\documentclass{svjour3}
\smartqed
\usepackage{graphicx}
\usepackage[sort&compress]{natbib}

\begin{document}

\title{A new simulation-based model for calculating post-mortem intervals using developmental data for \textit{Lucilia sericata} (Dipt.: Calliphoridae)}

\author{Saskia Reibe        \and
        Philip v. Doetinchem \and Burkhard Madea 
}

\institute{S. Reibe and B. Madea \at
              Institute for Forensic Medicine, University of Bonn, Stiftsplatz 12, 53111 Bonn \\
              Tel.: +49-179-5181634\\
              Fax: +49-228-738339\	
              \email{sreibe@googlemail.com}
           \and
           Ph. v. Doetinchem \at
              I. Physics Institute B, RWTH Aachen University, Sommerfeldstr.~14, 52074~Aachen
}

\date{Received: date / Accepted: date}

\maketitle

\begin{abstract}
Homicide investigations often depend on the determination of a minimum post-mortem interval (PMI$_{min}$) by forensic entomologists. The age of the most developed insect larvae (mostly blow fly larvae) gives reasonably reliable information about the minimum time a person has been dead. Methods such as isomegalen diagrams or ADH calculations can have problems in their reliability, so we established in this study a new growth model to calculate the larval age of \textit{Lucilia sericata} (Meigen 1826). This is based on the actual non-linear development of the blow fly and is designed to include uncertainties, e.g. for temperature values from the crime scene. We used published data for the development of \textit{L. sericata} to estimate non-linear functions describing the temperature dependent behavior of each developmental state. For the new model it is most important to determine the progress within one developmental state as correctly as possible since this affects the accuracy of the PMI estimation by up to 75\%. We found that PMI calculations based on one mean temperature value differ by up to 65\% from PMIs based on an 12-hourly time temperature profile. Differences of 2 $^\circ$C in the estimation of the crime scene temperature result in a deviation in PMI calculation of 15 - 30\%. 


\end{abstract}

\section{Introduction}
\label{intro}

Insect development rates are applied not only in pest control management but also in forensic entomology \citep{Greenberg1991}. Several species of dipteran and coleopteran families infest decaying material in order to breed offspring, and this includes the colonization of animal carcasses as well as dead bodies \citep{Lane1975, Archer2003a,Grassberger2004}. In homicide investigations, determination of the age of larvae feeding on a corpse can indicate a minimal post-mortem interval (PMI$_{min}$) \citep{Goff1993b}. This is often important in forensic case work \citep{Benecke1998a}. 

\subsection{Life cycle of blow flies}

The general life cycle of blow flies includes four stages: egg stage, larval stage, pupal stage and imago stage \citep{Tao1927}. During the larval stage, three instars can be separated: 1st, 2nd and 3rd instar,  where the latter is divided due to behavioral changes in feeding and post-feeding larvae. 

Blow flies deposit egg clutches directly on the food substrate, such as a dead body \citep{Smith1997}, in a position where the eggs are protected and in a moist environment. This ensures a food supply for the hatching 1st instar larvae. The first three instars each undergo a moult to reach the next developmental stage; the stages can be distinguished by the number of respiratory slits at the posterior end of the larvae. The third instar stage lasts for longer than the first two and is divided in a feeding and a post-feeding phase. The latter is a preparation for pupation. Therefore, the larvae leave the food source to find a suitable place for pupation, emptying their gut \citep{Arnott2008}. About one third of the pre-adult development time is spent in the post-feeding larval stage \citep{Greenberg1991}. Then pupation sets in and the imago develops within the pupal case till eclosion \citep{Lowne1890}. This last stage persists for about half of the time of the total development. 

The larval growth rate depends on its body temperature, which is directly influenced by environmental conditions as ambient temperature and the heat generated by maggot aggregations \citep{Slone2007}. Also, an important detail for PMI determination is that each species has its own temperature dependent growth rate.

\subsection{Methods for PMI determination}

In forensic case work, two different methods are frequently used to calculate a PMI. The first uses isomegalen  or isomorphen diagrams, by which the lengths or the developmental stage of the larvae are combined as a function of time and mean ambient temperature  in a single diagram \citep{Grassberger2001}. According to its originators, this method is optimal only if the body and therefore the larvae were not undergoing fluctuating temperatures, e.g. in an enclosed environment where the temperature was nearly constant. 

The second method of calculating a PMI estimates the Accumulated Degree Days or Hours (ADD or ADH). ADH values represent a certain number of "energy hours" that are necessary for the development of insect larvae. The degree day or hour concept assumes that the developmental rate is proportional to the temperature within a certain species-specific temperature range (overview in \citep{Higley2009}). However, the relationship of temperature and development rate (reciprocal of development time) is typically curvilinear at high and low temperatures and linear only in between.
 
The formula for calculating ADH is given by
\begin{equation}
\mbox{ADH} = T\cdot (\Theta - \Theta_0)
\label{e-adh}
\end{equation}

where $T$ is the development time, $\Theta$ is the ambient temperature, and the minimum developmental threshold temperature $\Theta_0$ is a species-specific value, the so called development zero, which is the x-intercept, i.e., an extrapolation of the linear approximation of the reciprocal of the developmental time. This value has no biological meaning, it is the mathematical consequence of using a linear regression analysis \citep{Higley2009}.

\begin{figure}
{\includegraphics[height=1.0\linewidth]{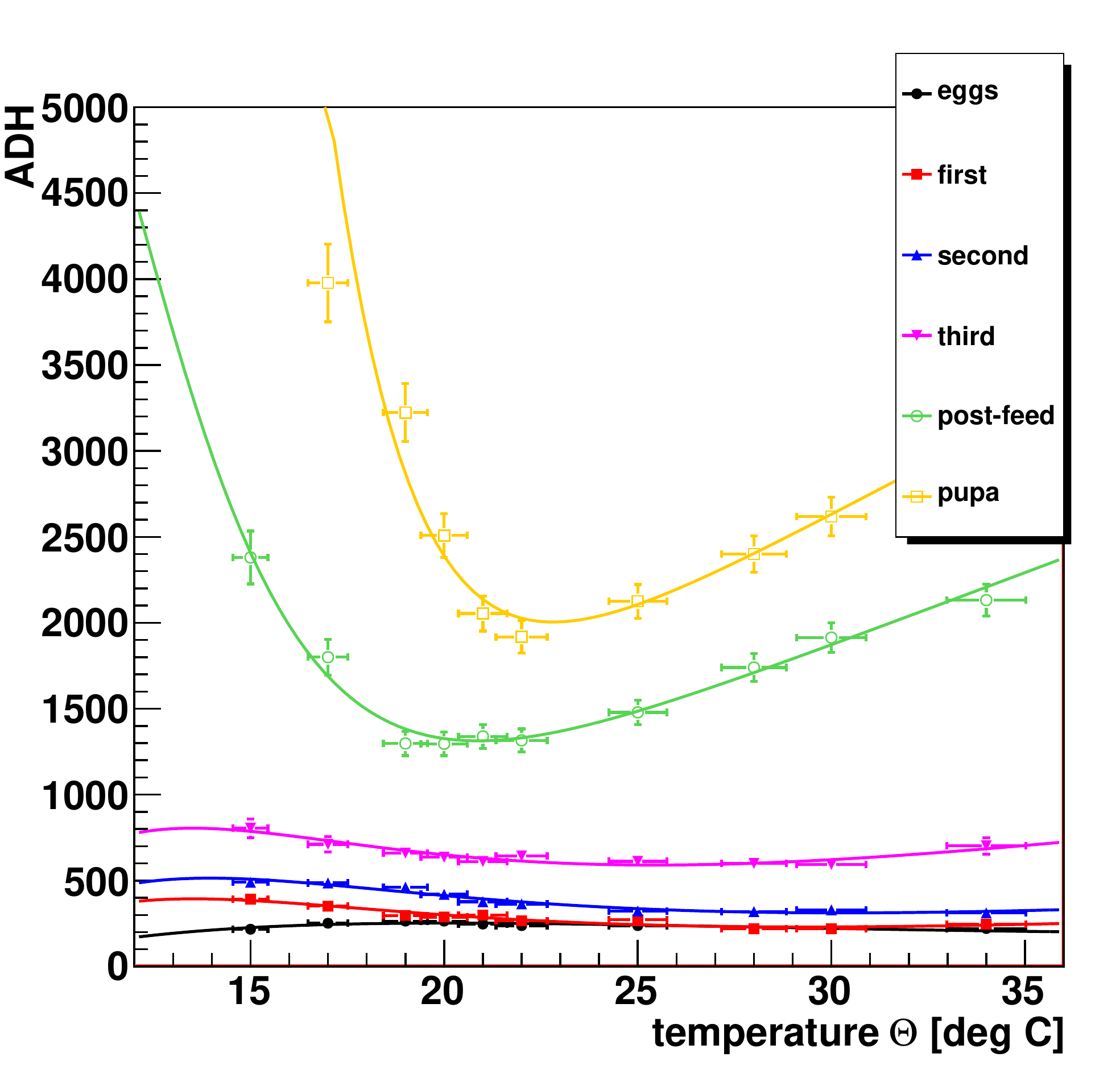}}
\caption{Calculated ADH values for the development of \textit{L. sericata} using eq. (\ref{e-adh}) (data points) and fitted functions (lines) calculated using eq. (\ref{e-fit}) and estimated parametres (Table \ref{t-fitpar}). Original data by Grassberger and Reiter 2001.}
\label{f-temp_adh}
\end{figure}

One basic condition for using the ADH method is that the ADH value for completing a developmental stage stays constant within certain temperature thresholds. For example a developmental duration for finishing a certain stage of 14 days at 25$^\circ$C results in 238 ADD when a base temperature of 8$^\circ$C is assumed. A developmental duration of 19 days at 21$^\circ$C results in 231 ADD, both ADD-values are in the same range.
We analyzed a published data-set for the development of \textit{Lucilia sericata} (Meigen 1826) \citep{Grassberger2001} and calculated the corresponding ADH values for these data.
Fig.~\ref{f-temp_adh} shows the calculated ADH values for a base temperature of $\Theta_0$=8$^\circ$C (as calculated by a linear regression analysis for the used data-set). 
In the figure we see a new effect: for the younger and also shorter developmental phases the ADH values are nearly constant over the complete range of temperatures, but for the post-feeding and the pupal stages the ADH values are strongly temperature dependent.

In general, the ADH method seems to give good results only when the larvae of interest have been exposed to temperatures similar to those used in generating the reference value applied in the PMI calculation \citep{Anderson2001}.  Moreover, the temperature range in which the development rate is actually linear is not wide enough to cover all temperatures during a typical summer in Germany (see also examples for May/June 2008 in Fig.~\ref{f-temp_profile}). Furthermore, neither developmental durations nor base temperatures for development have been calculated for species originating from Germany. The method must therefore be used carefully.

Furthermore, it is highly problematic that uncertainties for temperature measurements from a crime scene cannot be taken into account by either of the commonly used methods for PMI determination. It is difficult to determine the actual temperature controlling the larvae at a real crime scene. Since temperature is the variable that most influences development, it is crucial to consider it as accurately as possible. The standard procedure is to use temperatures of the nearest weather station for the desired time frame and correct them by applying a regression starting from temperatures measured at the crime scene, when taking the larvae as evidence \citep{Archer2004a}. The corrected values still contain uncertainties that cannot be accounted for by the methods currently used for PMI determination. No information exists for either model about the quality of the method or the error intervals of the calculated PMIs.

\section{New approach for PMI determination}

We analyzed developmental data for \textit{L. sericata} at different temperatures and fitted an individual exponential function for each developmental stage. Data used as input to the model were published by Grassberger and Reiter (2001) and represent the minimal time in hours to complete each larval phase (egg stage = stage 0, 1st instar = stage 1, 2nd instar = stage 2, 3rd instar feeding = stage 3, 3rd instar post-feeding = stage 4 and pupal stage = stage 5) until eclosion of the adult blow fly. The used data-set is one of the rare sets which covers a lot of temperatures and the resulting growth curve seems to represent growth behavior well (see original paper). Unfortunately, Grassberger and Reiter do not give any error values for their measurements, so we assumed an error for the developmental times of about 1 hour.
These authors used 250\,g of raw beef liver in plastic jars, and placed 100\,eggs on the food substrate. The jars were placed in a precision incubator. At each temperature regime the procedure was repeated 10\,times. Every 4\,hours, four of the most developed maggots were removed from the plastic jars, killed in boiling water, and preserved in alcohol \citep{Adams2003} and then their stage of development was determined.

\subsection{Data fit}

\begin{figure}
{\includegraphics[height=1.0\linewidth]{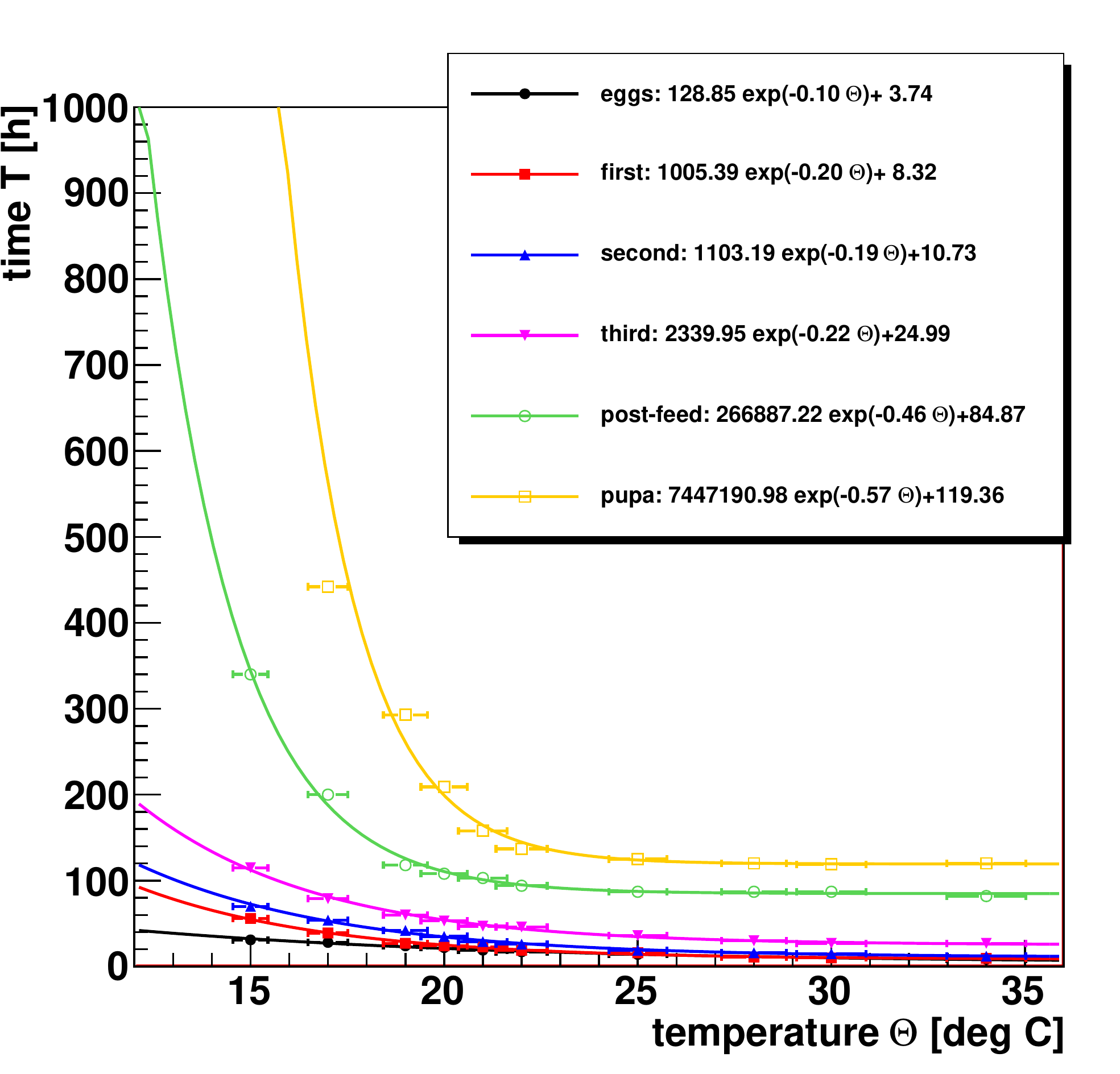}}
\caption{Developmental data of \textit{L. sericata} with fitted functions (eq. (\ref{e-fit})). Original data by Grassberger and Reiter 2001.}
\label{f-temp_time}
\end{figure}

Our new larval growth model is based on the data shown in Fig.~\ref{f-temp_time}, in which the duration of each developmental stage was measured as a function of temperature \citep{Grassberger2001}. These data points were fitted with an exponential function of the form:

\begin{equation}
T_\alpha(\Theta)=a_\alpha\cdot\exp\left(-\tau_\alpha\cdot\Theta\right)+T_{0,\alpha}
\label{e-fit}
\end{equation}

where $T_\alpha$ is the duration of one developmental stage $\alpha$ as a function of temperature $\Theta$. The parameters fitted for the different stages are shown in Table~\ref{t-fitpar}. The parameter $\tau_\alpha$ defines how strongly the time interval depends on temperature; the higher the parameter in Table~\ref{t-fitpar}, the steeper is the gradient of the fitted curve. $T_{0,\alpha}$ represents the minimum time interval required for finishing a certain developmental stage and $a_\alpha$ provides the absolute normalization. The developmental stages of the maggots were determined every $\Delta T = 4$\,h, such that time measurement errors are set to $\sigma_T=\Delta T/\sqrt{12}$ following an uniform distribution. It is assumed that the maggot body temperature is known to an accuracy of 3\,\% in order to take into account uncertainties about differences between ambient and maggot body temperature. The parameters $a_\alpha$, $\tau_\alpha$ and $T_{0,\alpha}$ were determined by minimizing the sum of  error squares. As seen in Fig.~\ref{f-temp_time}, the exponential function accurately models the behavior during all developmental stages and will be used below.  In all stages the developmental duration at temperatures below 24 $^\circ$C starts to rise exponentially. 
Fig.~\ref{f-temp_adh} shows the  calculated ADH values corresponding to eq.~(\ref{e-adh}) (data points). In addition, the figure shows the function $\mbox{ADH}_\alpha(\Theta)=T_\alpha(\Theta)\cdot(\Theta-\Theta_0)$ (lines). $T_\alpha(\Theta)$ is calculated by eq.~(\ref{e-fit}) with the previously fitted parameters (Table~\ref{t-fitpar}). Again, the functions give a reasonable description of the data. Nevertheless, the model is an empirical one, based on the observations of the data points generated by Grassberger and Reiter (2001).

\subsection{PMI calculation}

For European and especially German temperatures, calculation of the total developmental duration must allow for non-linear temperature behavior in order to ensure accuracy. The basic idea underlying a new approach in PMI determination is to follow an ambient time-temperature profile $\Theta(t)$ backwards in time starting from the time point $t_F$ at which the maggots of interest were collected. The idea of backwards calculation is obviously similar to the ADH method, but in the new model the important improvement is the way of calculating  the larval age. The latter is calculated successively during certain time steps using the fitted functions (introduced in Fig.~\ref{f-temp_time}) corresponding to the current developmental stage. In each stage $\alpha$ the relative developmental progress is $P_\alpha$ (values 0 - 1) where 0 is the beginning and 1 is the finishing point of each developmental stage; e.g. a maggot in the middle of the post-feeding stage is $P_4 = 0.5$, at the end of the post-feeding stage it is $P_4 = 0.9$ and so forth.  The developmental duration $t_{\alpha,0}$ spent in each individual stage is calculated by solving the relation:
\begin{equation}
P_\alpha = \int_{t_{\alpha+1,0}}^{t_{\alpha,0}} \frac{\mbox{d}t}{T_\alpha(\Theta(t_F-t))}\label{e-prog}.
\end{equation}

where $dt/T(\Theta( t ))$ is the infinitesimal relative development.

The calculation starts with the developmental stage of the maggot at the time of collection,  summing the developmental progress of each stage backwards until the beginning of the egg stage is reached. The calculation for each collection stage uses $t_{\alpha+1,0} = t_F$. The total development time $t_0$  or post-mortem interval (PMI) is then given by \begin{equation}t_0 = \sum_\alpha t_{\alpha,0}.\end{equation}

For the new model a program was written in C++ using Root (http://root.cern.ch/). This program includes all mentioned mathematical steps and produces the figures shown here as output. For each new PMI calculation, the corresponding temperature profile can be inserted and individually chosen uncertainties can be included.

\subsection{Consideration of uncertainties by Monte-Carlo simulation}

To explore the uncertainties in the total developmental duration, a Monte-Carlo simulation was applied, which is commonly used for simulations in life sciences \citep{Mansson2005}. It is a method for calculating one final uncertainty after considering all statistically independent uncertainties that influence e.g. the larval age. The mean PMI with corresponding standard variation is calculated $n$ times taking into account and varying all uncertainties described in the following.  First, the developmental profiles $T_\alpha(\Theta)$ have uncertainties due to the measurement procedure. Second, the time-temperature profile from the collection scene is not known precisely and must be approximated using temperature values from nearby weather stations. The variations are introduced for each model as follows:
\begin{description}
\item[Development profile:] The mean duration values of the temperature-time data are randomly smeared with a uniform distribution with corresponding error $\sigma_{T}$; for the maggot body temperature $\sigma_{\Theta,b}$ a Gaussian distribution is used. New fits with the function in eq.~(\ref{e-fit}) are performed for each stage.
\item[Time-temperature profile:] Deviations between the temperature profile at the collection scene and the nearest weather station are accounted for by Gaussian smearing of time $t$ and temperature $\Theta$, with the corresponding errors $\sigma_{t}$ and $\sigma_\Theta$ as width for each data point. $\sigma_\Theta$ can be inserted in the model`s calculation individually dependent on the differences between the temperatures at finding place and weather station.
\end{description}

\begin{figure}
\rotatebox{90}{\includegraphics[height=1.0\linewidth]{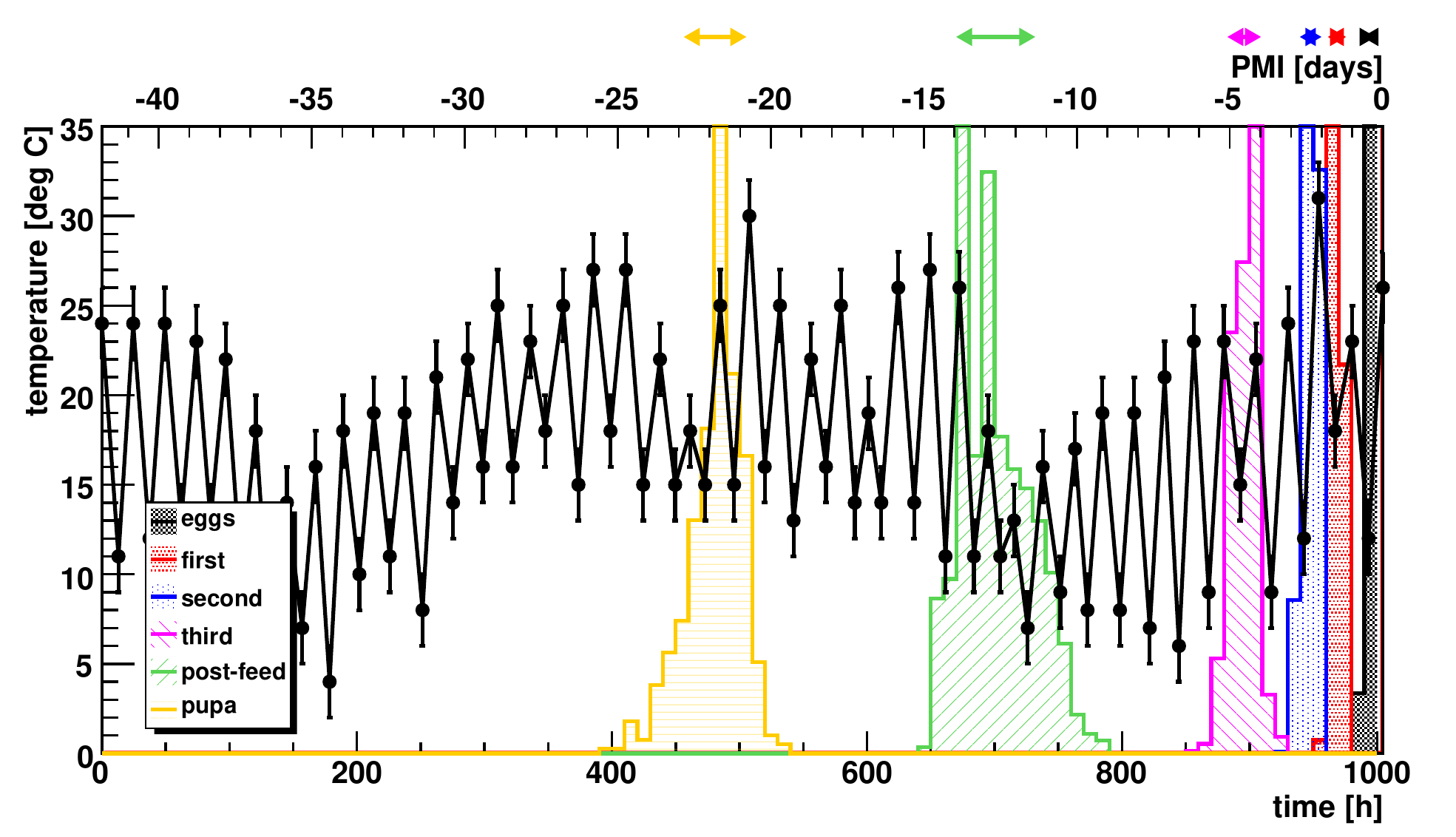}}
\rotatebox{90}{\includegraphics[height=1.0\linewidth]{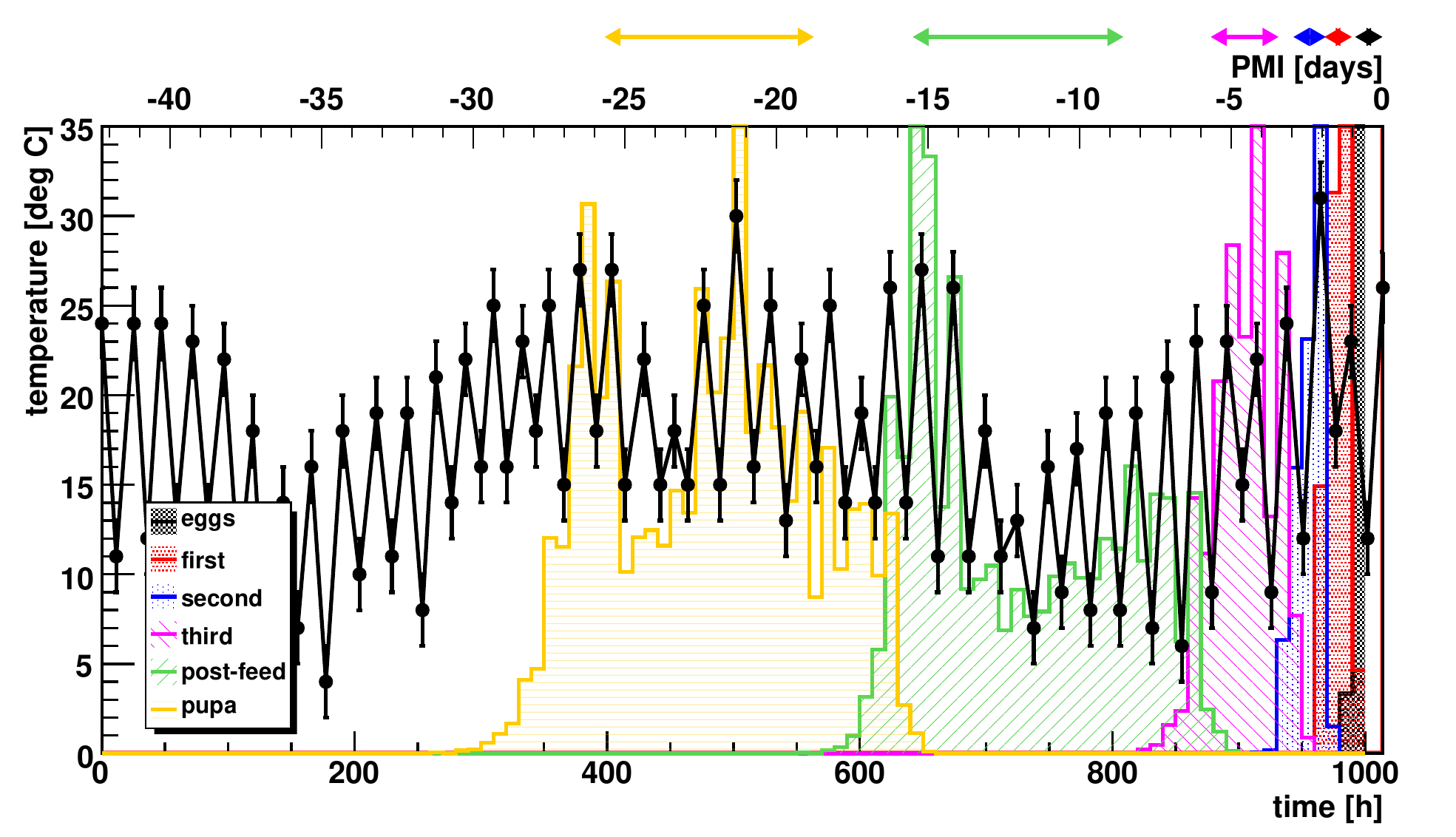}}
\caption{Excerpt of the temperature profile during May and June 2008 in Cologne/Bonn, Germany (www.wetteronline.de). The histograms illustrate the PMI distributions of the different random models for a certain collection stage $\alpha$. The arrows at top show 1-standard deviation intervals around the mean PMIs for each stage. \textit{\textbf{Upper part:}} PMI calculation with parameters $\sigma_{T}=4/\sqrt{12}$\,h, $\sigma_{t}=1$\,h, $\sigma_\Theta=2^\circ$C, $\sigma_{\Theta,b} = 3$\,\% for 10.000 models for a fixed collection stage progress of $P_\alpha =0.5$. \textit{\textbf{Lower part:}} PMI calculation as before but the progression of the stage is not fixed.}
\label{f-temp_profile}
\end{figure}

We calculated the PMI for a mock crime scene with the following parameters: the error of the measurement of the original data $\sigma_{T}=4/\sqrt{12}$\,h, the errors of the data of the weather station $\sigma_{t}=1$\,h and $\sigma_\Theta=2^\circ$C, the difference of the ambient temperature and larval body temperature $\sigma_{\Theta,b} = 3$\,\% for 10.000 models for a fixed collection stage progress of $P_\alpha =0.5$. The results are shown in Fig.~\ref{f-temp_profile}~\textit{(upper part)} which is a direct output of the new program that calculates the PMI.  The lower time axis defines the progress of the temperature profile forward in time, representing the time frame of interest. The temperature profile used here (black line) is taken from the minimum and maximum temperatures in May and June 2008 measured at Cologne/Bonn airport. The right end of the diagram marks a fictional time point of maggot collection and therefore the starting point for PMI calculation. The upper time axis depicts the PMI backwards in time starting from the moment of maggot collection. For each developmental stage the PMI was calculated by following a linear interpolation between the maximum and minimum temperatures. The histograms illustrate the PMI distribution for each stage and show a clear single peak structure. The arrows on the top show the 1-standard deviation interval for each stage around the mean PMI value, and range between 0.1 and 1.2 days (depending on the stage).
Since no data points below temperatures $\Theta < 15^\circ$C were measured, the functions $T_\alpha(\Theta)$ were extrapolated to lower temperatures. As expected, the PMI and the corresponding standard deviation increase with higher developmental duration (see arrows above histogram). 

Since the exact progress within the developmental stage at collection time is most of the time also unknown, a third uncertainty is introduced:

\begin{description}
\item[Stage progress:] The developmental stage at collection time was determined only to integer precision, so that it is assumed the exact progress is an uniformly distributed value between 0 and 1. Consequently, the starting value for the PMI calculation $P_\alpha$ at time $t_F$ is randomly and uniformly chosen within the interval~[0,1] for each model.
\end{description}

Fig.~\ref{f-temp_profile}~\textit{(lower part)} shows the PMI calculation for the same parameters as before, but without setting the progress of the development for each stage to a fixed value. The 1-standard deviation values increase by 0.3 to 3.3 days. The resulting uncertainty in the progress of the stage contributes about 75\,\% to the total PMI error interval. In addition, the histograms show deviations from a clear single peak structure, e.g. for the pupal stage, implying that the PMI probabilities for 21\,days and 26\,days are nearly the same. To use the new model, the crucial parameter is therefore the correct determination of the progress of the developmental stage of maggots collected from a corpse.

\subsection{Estimation of temperature at the location of maggot collection}

\begin{figure}
\rotatebox{90}{\includegraphics[height=1.0\linewidth]{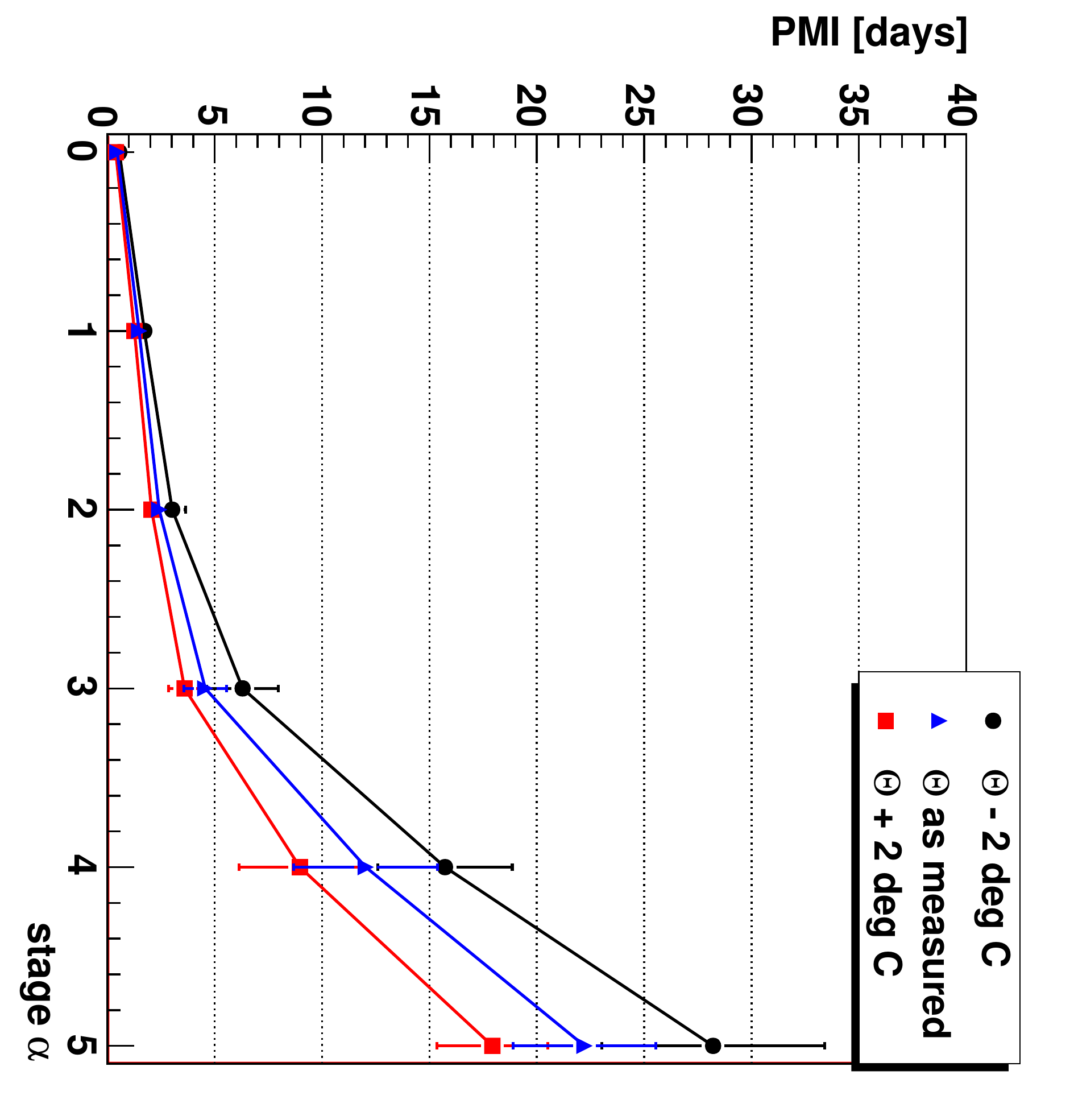}}
\caption{Calculated PMIs and SDs for each stage for a temperature profile of May/June 2008 measured at Cologne/Bonn airport in comparison to profiles $\pm$ 2 $^\circ$C. The later the stage the greater the SD and the differences in the calculated PMI for differences of 2 $^\circ$C.}
\label{f-stage_pmi_compare}
\end{figure}

The impact of correct temperature determination at the maggot collection scene is shown in Fig.~\ref{f-stage_pmi_compare}. The data points represent the mean PMIs with an error bar of 1 standard deviation as a function of collection stage for three different temperature profiles. The triangles show the PMIs for the original temperature profile as measured at Cologne/Bonn airport. The bullets (squares) show the results for the same profile but subtracted (and added) by 2 $^\circ$C. As expected, the PMIs and the corresponding standard deviations of the lower (higher) temperature profile increase (decrease) relative to the nominal profile. These differences in temperature of 2 $^\circ$C give rise to an effect of 15 - 30\,\%. That implies that a miscalculation of the temperature at the crime scene of 2 $^\circ$C will result in a miscalculation of the PMI by 15-30\,\%. The later the stage, the greater the deviation from the actual PMI.

\subsection{Comparing PMIs based on mean temperatures and 12-hourly temperature profiles}

\begin{figure}
\rotatebox{90}{\includegraphics[height=1.0\linewidth]{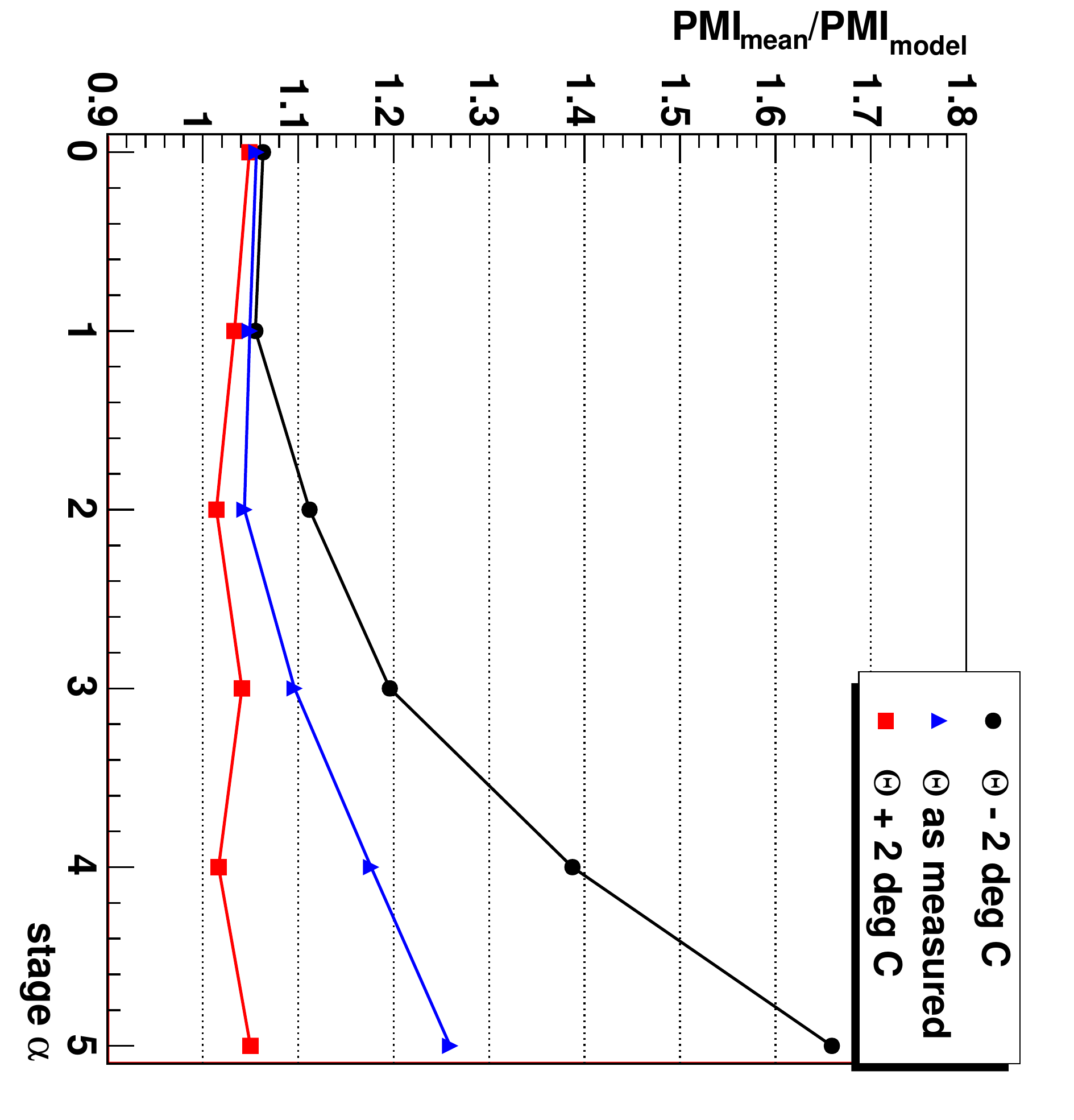}}
\caption{Deviations of calculated PMIs when using the mean temperature and the actual 12-hourly temperature profile of May/June 2008 from Cologne/Bonn airport. Additionally, the results are shown for the temperature profiles $\pm$ 2 $^\circ$C.}
\label{f-pmi_rel_compare}
\end{figure}

In Fig.~\ref{f-pmi_rel_compare}, PMIs calculated using the corresponding mean temperature values in the temperature interval $[t_0,t_F]$ are compared with PMIs from our model using the three temperature profiles introduced previously. The calculated mean temperatures were as follows (calculated for the time frame till completion of each stage): stage 0 = 18$^\circ$C, stage 1 = 19 $^\circ$C, stage 2 = 20 $^\circ$C, stage 3 = 19 $^\circ$C, stage 4 = 16 $^\circ$C, stage 5 = 17 $^\circ$C. The PMI values based on the temperature profile and those based on a mean temperature value agree to within about 5\,\% for the high temperature value (original profile + 2 $^\circ$C) in all  stages. The deviation between mean temperature and the original temperature profile exceeds the 10\,\% level starting at the 3rd instar feeding stage, and increases to 25\,\% in the pupal stage. This effect becomes even larger for the low temperature profile (original profile subtracted by 2 $^\circ$C). Starting from the 2nd instar stage, the deviation increases from about 10\,\% up to about 65\,\% for the pupal stage. This means that use of mean temperature values overestimates the influence of low temperatures and underestimates periods of high temperatures. The effect should be larger if the mean temperature during the development is lower still, e.g. in spring or fall. In general, more data points are needed for the developmental duration at low temperature ranges to provide more reliable statements.

\subsection{Does the model work in a real case?}

We calculated the PMI in a real case where the actual PMI was known due to a confession of the offender. At the end of August 2007 the victim was killed in early morning and was found 4 days later also in the morning on a grassland. This leads to a PMI of approximately 96 hours. The victim was stabbed to death and had several wounds which would act as attractant to the blow flies. It can be assumed that blow flies started ovipositing early after death occurred \citep{Reibe2010}. Autopsy was performed directly after the corpse was recovered and several 2nd instar larvae of \textit{L. sericata} were collected. The largest larvae measured 6.1 mm. Hourly temperature values were taken from a weather station 10 km away. The mean temperature was 16 $^\circ$C. Using Grassberger and Reiter`s isomegalen diagram for a larvae measuring 6 mm and a mean temperature of 16~$^\circ$C results in a time interval of 3.2 days plus 30 hours (larval development time plus egg period). In total a PMI of 107 hours is indicated. This would shift the time of oviposition to nighttime, which is a highly unlikely event \citep{Amendt2008}. The same data can be used to calculate the ADH value for \textit{L. sericata} for reaching 6 mm in order to calculate the PMI not based on the mean temperature but on hourly data. As mentioned earlier, a regression analysis of the data set reveals a base temperature of 8 $^\circ$C. The corresponding ADH value is therefore 856, based on the equation:

\begin{equation}
\mbox{ADH} = 107 (16 - 8)
\label{e-adh_case}
\end{equation}

Subtracting the hourly ADH values, estimated by the temperature values from the weather station and the base temperature, from the starting value of 856 results in a PMI of 101 hours.

To use the new model for calculating larval age in the real case, information about the progress of the 2nd instar larval stage was required. In the original work of Grassberger and Reiter (2001, Figure 1) a figure is included showing the growth of the larvae and also the time points for each moult. According to this figure, the 2nd instar stage sets in after the larvae have reached a size of approximately 4 mm and ends when the larvae have reached a size of approximately 8mm. As the largest larvae we collected measured 6 mm, we chose P=0.5 as progress for the larval stage. We included the hourly temperature profile and chose a temperature error of 1 $^\circ$C. The result of the calculation was a PMI of 99 hours (SD = 3 hours). 

These calculations of a PMI in a real case show that all three methods give reasonable results. Furthermore, it becomes obvious that the new model is a possible alternative for the existing methods with the benefit of directly providing a standard deviation for the calculation. 

\section{Conclusion and outlook}

The new model improves the larval age calculation in specific ways. It can be used in non-linear parts of the temperature dependent development, and includes individually defined uncertainties for a temperature profile determined retrospectively from the nearest weather station. In the new model the temperature profile plus the determination of the larval stage are translated into a mean PMI as well as a standard deviation. PMI calculation using mean temperatures, however, can lead to severe deviations from the real PMI. 

So far, the main uncertainty arises from the fact that the developmental stage is determined only on a 1 - 6 scale (egg, 1st instar, 2nd instar, 3rd instar feeding, 3rd instar post-feeding and pupae). As shown above, 75\,\% of the uncertainties in the model depend on the exact determination of the developmental progress, and additional length values, as shown for the PMI calculation in the real case, will propably increase its accuracy leading to more accurate PMI calculations. Moreover, the next step is to produce own growth data with known error values to refine the inclusion of uncertainties that are only rough estimates at the present time and to improve the till now only empirical model.

Nevertheless, the new PMI calculation program is suitable for use in forensic case work as a general tool for PMI determination. Scientists from every country or climatic region can incorporate their own growth values for different species and ensure a high accuracy in PMI determination.

\pagebreak

\begin{table}
\begin{center}
\caption{Fitted parameters for the development-time-function $T_\alpha(\Theta)$.}
\begin{tabular}{cl|ccc}
\hline
\hline
$\alpha$	& Stage		& $a_\alpha$ [h]	& $\tau_\alpha$ [$^\circ$C$^{-1}$]		& $T_{0,\alpha}$ [h]\\
\hline
0		& eggs		& $1.28\cdot10^{2}$			& 0.10						& 3.74\\
1		& 1st			& $1.00\cdot10^{3}$			& 0.20						& 8.32\\
2		& 2nd			& $1.10\cdot10^{3}$			& 0.19						& 10.73\\
3		& 3rd			& $2.34\cdot10^{3}$			& 0.22						& 24.99\\
4		& post-feed	& $2.67\cdot10^{5}$			& 0.46						& 84.87\\ 
5		& pupa		& $7.45\cdot10^{6}$			& 0.57				 		& 119.36\\
\hline
\end{tabular}
\label{t-fitpar}
\end{center}
\end{table}

\begin{thebibliography}{19}
\providecommand{\natexlab}[1]{#1}

\bibitem[{Adams and Hall(2003)}]{Adams2003}
Adams Z and Hall M (2003) Methods used for the killing and preservation of
  blowfly larvae, and their effect on post-mortem larval length.
\newblock Forensic Sci Int 138(1-3):50--61

\bibitem[{Amendt et~al.(2008)Amendt, Zehner and Reckel}]{Amendt2008}
Amendt J, Zehner R and Reckel F (2008) The nocturnal oviposition behaviour of
  blowflies ({D}iptera: {C}alliphoridae) in {C}entral {E}urope and its forensic
  implications.
\newblock Forensic Sci Int 175(1):61--64

\bibitem[{Anderson(2001)}]{Anderson2001}
Anderson G (2001) \emph{Forensic Entomology: The Utility of Arthropods in Legal
  Investigations}, chapter Insect Succession on Carrion and its Relationship to
  Determining Time of Death.
\newblock CRC Press, pp. 143--175

\bibitem[{Archer and Elger(2003)}]{Archer2003a}
Archer M and Elger M (2003) Female breeding-site preferences and larval feeding
  strategies of carrion-breeding Calliphoridae and Sarcophagidae (Diptera): a
  quantitative analysis.
\newblock Australian Journal of Zoology 51:165--174

\bibitem[{Archer(2004)}]{Archer2004a}
Archer MS (2004) The effect of time after body discovery on the accuracy of
  retrospective weather station ambient temperature corrections in forensic
  entomology.
\newblock J Forensic Sci 49(3):553--559

\bibitem[{Arnott and Turner(2008)}]{Arnott2008}
Arnott S and Turner B (2008) Post-feeding larval behaviour in the blowfly,
  {C}alliphora vicina: effects on post-mortem interval estimates.
\newblock Forensic Sci Int 177(2-3):162--167

\bibitem[{Benecke(1998)}]{Benecke1998a}
Benecke M (1998) Six forensic entomology cases: description and commentary.
\newblock J Forensic Sci 43(4):797--805; 1303

\bibitem[{Goff(1993)}]{Goff1993b}
Goff M (1993) Estimation of Postmortem Interval Using Arthropod Development and
  Successional Patterns.
\newblock Forensic Science Review 5 (2):81--94

\bibitem[{Grassberger and Frank(2004)}]{Grassberger2004}
Grassberger M and Frank C (2004) Initial study of arthropod succession on pig
  carrion in a central {E}uropean urban habitat.
\newblock J Med Entomol 41(3):511--523

\bibitem[{Grassberger and Reiter(2001)}]{Grassberger2001}
Grassberger M and Reiter C (2001) Effect of temperature on {L}ucilia sericata
  ({D}iptera: {C}alliphoridae) development with special reference to the
  isomegalen- and isomorphen-diagram.
\newblock Forensic Sci Int 120(1-2):32--36

\bibitem[{Greenberg(1991)}]{Greenberg1991}
Greenberg B (1991) Flies as forensic indicators.
\newblock J Med Entomol 28(5):565--577

\bibitem[{Higley and Haskell(2009)}]{Higley2009}
Higley L and Haskell N (2009) \emph{Forensic Entomology, The Utility of
  Arthropods in Legal Investigations 2nd ed.}, chapter Insect Development and
  Forensic Entomology.
\newblock CRC Press LLC, pp. 389--405

\bibitem[{Lane(1975)}]{Lane1975}
Lane RP (1975) An investigation into blowfly (Diptera: Calliphoridae)
  succession on corpses.
\newblock J Nat Hist 9:581--588

\bibitem[{Lowne(1890)}]{Lowne1890}
Lowne B (1890) \emph{The anatomy, physiology, morphology and development of the
  blow-fly.}
\newblock RH Porter, London

\bibitem[{Mansson et~al.(2005)Mansson, Frey, Essex and Welsh}]{Mansson2005}
Mansson RA, Frey JG, Essex JW and Welsh AH (2005) Prediction of properties from
  simulations: a re-examination with modern statistical methods.
\newblock J Chem Inf Model 45(6):1791--1803

\bibitem[{Reibe and Madea(2010)}]{Reibe2010}
Reibe S and Madea B (2010) How promptly do blow flies colonise fresh carcasses?
  A study comparing indoor vs. outdoor locations.
\newblock For Sci Int 195:52--57

\bibitem[{Slone and Gruner(2007)}]{Slone2007}
Slone D and Gruner S (2007) Thermoregulation in larval aggregations of
  carrion-feeding blow flies (Diptera: Calliphoridae).
\newblock J Med Entomol 44(3):516--523

\bibitem[{Smith and Wall(1997)}]{Smith1997}
Smith KE and Wall R (1997) The use of carrion as breeding sites by the blowfly
  {L}ucilia sericata and other {C}alliphoridae.
\newblock Med Vet Entomol 11(1):38--44

\bibitem[{Tao(1927)}]{Tao1927}
Tao S (1927) A Comparative Study of the early Larval Stages of some common
  Flies.
\newblock Am J Epidemiol 7:735 -- 761

\end{thebibliography}
\end{document}